\documentclass[11pt]{article}

\usepackage[margin=1in]{geometry}
\usepackage[T1]{fontenc}
\usepackage{times}
\usepackage[hyphens]{url}
\usepackage{graphicx}
\usepackage[round,authoryear]{natbib}
\usepackage[font=small,labelfont=bf]{caption}
\usepackage{algorithm}
\usepackage{algorithmic}
\usepackage{booktabs}
\usepackage{amsmath,amssymb}
\usepackage{tikz}
\usepackage{pgfplots}
\usepgfplotslibrary{groupplots,fillbetween}
\pgfplotsset{compat=1.18}
\usepackage{microtype}
\usepackage{xcolor}
\usepackage[hidelinks]{hyperref}

\definecolor{mappoblue}{RGB}{31,119,180}
\definecolor{eventorange}{RGB}{255,127,14}
\definecolor{tocagreen}{RGB}{44,160,44}
\definecolor{ablred}{RGB}{214,39,40}

\title{Turnover-Orthogonal Credit Assignment for Open-Team Multi-Agent Reinforcement Learning}
\author{%
Amit Thakur \qquad Mukesh Singhal\\[0.35em]
\small University of California, Merced
}
\date{}

\hypersetup{
  pdftitle={Turnover-Orthogonal Credit Assignment for Open-Team Multi-Agent Reinforcement Learning},
  pdfauthor={Amit Thakur and Mukesh Singhal},
  pdfkeywords={multi-agent reinforcement learning, open teams, credit assignment, turnover, permutation invariance}
}

\begin{document}

\maketitle

\begin{abstract}
Open-team multi-agent reinforcement learning studies cooperative systems in which agents may join, leave, or be replaced during an episode. In such settings, the team return changes both because agents choose useful actions and because the active population itself changes. Standard centralized critics and shared advantages often mix these two effects into one scalar credit signal, allowing surviving agents to be rewarded or penalized for exogenous turnover events outside their control. We introduce turnover-orthogonal credit assignment (TOCA), a value decomposition for open teams that separates action effects, pure turnover effects, and action--turnover interactions. Under exogenous turnover, the event-conditioned value admits a centered decomposition whose event-conditioned baseline removes the pure turnover component while preserving credit for actions that make the team robust to future replacements. We instantiate this idea with a permutation-invariant centralized critic over variable-size agent sets and event tokens, and derive both a counterfactual per-agent credit signal and a softly weighted interaction variant, TOCA-$\beta$, for high-variance control environments. Controlled diagnostic experiments show that TOCA improves return over event-aware MAPPO-style critics and that removing interaction credit substantially hurts performance. In a replacement-only Dynamic Spread benchmark, TOCA-$\beta$ achieves the best mean return at high turnover rates and improves over its no-interaction ablation. These results suggest that explicitly separating turnover from action credit is a useful principle for robust learning in dynamic cooperative teams.
\end{abstract}

\section{Introduction}

Cooperative multi-agent reinforcement learning (MARL) has made substantial progress in fixed-population settings, where the same agents interact throughout training and evaluation. Many real systems are not fixed: Robot teams lose units and receive replacements; ride-sharing and delivery fleets change membership throughout a shift; distributed sensing systems operate with intermittent devices; and human--AI teams may change participants mid-task. In these settings, agents must act while the active population is itself a dynamic part of the environment.

This paper studies credit assignment in such \emph{open-team} MARL. The core difficulty is that the learning signal conflates two different causes of value change: what the agents did, and what happened to the team. After a joining, leaving, or replacement event, a standard temporal-difference error can penalize a survivor because a strong teammate left, or reward it because a capable teammate arrived, even when that survivor repeated exactly the same action. Conversely, if an agent anticipates turnover and moves into a configuration that is robust to teammate loss, that benefit should remain part of its credit. A useful signal must therefore remove the \emph{pure} effect of turnover while preserving \emph{action-dependent robustness} to turnover.

This distinction is important both scientifically and practically: Scientifically, confounded credit makes it hard to tell whether a learned policy is genuinely robust or is merely benefiting from favorable team changes. Practically, misattribution can slow learning, destabilize coordination, and produce brittle behavior in domains where team membership changes are routine. In robotics, logistics, sensing, and human--AI collaboration, the desired behavior is not just high average return, but resilience under teammate loss, failure, or replacement. That makes source-of-credit separation a central problem rather than a cosmetic refinement.

Existing centralized training methods partially address coordination by conditioning critics on joint observations and actions \cite{lowe2017maddpg,foerster2018coma,yu2022mappo}. Value factorization methods decompose team value across agents \cite{sunehag2018vdn,rashid2018qmix,son2019qtran}. However, these methods are usually formulated for a fixed set of agents. Event-aware critics can condition on turnover information, but conditioning alone does not determine how much of the resulting value should be attributed to actions versus the event itself. We therefore argue that open-team credit assignment requires a decomposition with explicit orthogonality to pure turnover effects.

We propose \emph{turnover-orthogonal credit assignment} (TOCA). Let $h$ denote the centralized history, $a$ the joint action of active agents, and $e$ the turnover event. TOCA decomposes the event-conditioned value into
\begin{equation}
Q^{\pi}(h,a,e)=\mu(h)+q_A(h,a)+q_E(h,e)+q_{AE}(h,a,e),
\label{eq:decomp}
\end{equation}
where $\mu(h)$ is the average state value, $q_A$ is the action main effect, $q_E$ is the pure turnover effect, and $q_{AE}$ is the action--turnover interaction. The baseline $\mu(h)+q_E(h,e)$ is event-conditioned but action-independent. Subtracting it gives a turnover-orthogonal advantage that removes pure event credit while retaining $q_A+q_{AE}$. In high-variance control tasks, we also use TOCA-$\beta$, which softly weights the interaction term.

Our contributions are as follows: First, we formulate open-team credit assignment as a decomposition problem over actions and turnover events. Second, we derive an event-conditioned baseline that is orthogonal to pure turnover effects under exogenous turnover. Third, we instantiate the decomposition with a permutation-invariant critic suitable for variable-size teams. Fourth, we evaluate the method in a controlled diagnostic benchmark and a replacement-only Dynamic Spread benchmark, showing that interaction credit is important and that TOCA-$\beta$ improves high-turnover robustness.

\section{Related Work}

\textbf{Credit assignment in cooperative MARL:} Difference rewards and counterfactual baselines aim to isolate an agent's contribution to a team return. COMA marginalizes one agent's action with a centralized critic \cite{foerster2018coma}; MADDPG uses centralized critics for decentralized policies \cite{lowe2017maddpg}; and MAPPO, HATRPO/HAPPO, and recent practice studies show the strength of PPO-style and trust-region training in cooperative MARL \cite{schulman2017ppo,yu2022mappo,kuba2021happo,fu2022revisiting}. These methods improve action credit but do not explicitly separate action credit from population-turnover credit.

\textbf{Value factorization and sequence models:} VDN, QMIX, QTRAN, and related methods decompose joint value functions into agent-level terms with structural constraints \cite{sunehag2018vdn,rashid2018qmix,son2019qtran}. Recent transformer-based MARL models, including MAT and TransMix, provide stronger sequence or mixing architectures for cooperative decision making \cite{wen2022mat,khan2022transmix}. TOCA is complementary: it decomposes value by sources of variation--actions, turnover events, and their interaction--rather than only by agents or sequence position.

\textbf{Open, variable-size, and permutation-invariant teams:} Open-team problems require representations that do not depend on a fixed ordering or fixed number of agents. Deep Sets and Set Transformers provide permutation-invariant or equivariant neural architectures for set-valued inputs \cite{zaheer2017deepsets,lee2019set}. Recent work on dynamic team composition and open-environment cooperative MARL highlights settings where agents with different capabilities may join or leave during operation \cite{liu2021coach,yuan2023openmarl}. We build on this motivation by encoding the active agent set and turnover-event tokens with shared encoders. PettingZoo and multi-agent particle environments provide common tools for evaluation \cite{mordatch2018emergence,terry2021pettingzoo}; our Dynamic Spread benchmark extends this style of task with fixed-size replacement events.

\subsection{Why Event Conditioning Alone Is Not Enough}

A natural baseline for open teams is to append event information to a centralized critic. This can improve value prediction because the critic knows whether an unfavorable replacement occurred. However, better prediction alone does not guarantee better credit assignment. If the actor still updates from a target that mixes action-caused and event-caused variation, then the policy can be pushed by return changes that were independent of the sampled action. The issue is especially visible after disruptive replacement events: the critic predicts a lower return, but without an event-conditioned baseline the replacement shock can still be distributed across all survivors.

This is the sense in which TOCA is ``turnover-orthogonal.'' The actor signal should be orthogonal to value variation caused only by exogenous population change, while still preserving action--turnover interactions. A method that removed all event dependence would also erase robustness credit; TOCA removes only the action-independent part. This is also why TOCA is different from standard value factorization: factorization composes team value from agents, whereas TOCA decomposes team value into sources of variation relevant for credit assignment.
\section{Open-Team Problem Setting}

We consider a cooperative partially observable Markov game \cite{littman1994markovgames,oliehoek2016dec} with a dynamic active set $\mathcal N_t$. At time $t$, active agents observe local observations $o_t^i$, choose actions $a_t^i$, and receive a shared reward $r_t$. The active population evolves through turnover events. Let
\begin{equation}
\mathcal N_{t+1}=(\mathcal N_t\setminus \mathcal L_t)\cup \mathcal J_t,
\end{equation}
where $\mathcal L_t$ is the set of leaving agents and $\mathcal J_t$ is the set of joining agents. We denote the event by
\begin{equation}
E_t=e_t=(\mathcal J_t,\mathcal L_t,\kappa_t),
\end{equation}
where $\kappa_t$ stores event attributes such as replacement type, capability, or inherited position. In the neural critic, this structured event is encoded as an \emph{event token}: a compact vector containing the event type, joining/leaving masks, and attributes such as replacement capability or inherited-position flags.

Let $H_t=h$ be the centralized history available during training and let $A_t=a$ be the joint action of the currently active agents. The event-conditioned value is
\begin{equation}
Q^{\pi}(h,a,e)=\mathbb E_{\pi}[G_t\mid H_t=h,A_t=a,E_t=e]
\end{equation}
where $G_t=\sum_{k\geq 0}\gamma^k r_{t+k}$ is the discounted return. We assume decentralized execution: each agent acts from its local observation and identity-free features, while the critic may use centralized information during training.

\subsection{Illustrative Open-Team Credit Example}

Consider a navigation team with two landmarks and two active agents. If a fast agent is replaced by a slow agent, the team return may immediately drop even if the remaining agent repeats the same action. This drop is a pure turnover effect. Now suppose the remaining agent moves toward the landmark previously covered by the fast agent. The usefulness of this action depends on the replacement event: it may be unnecessary if no replacement happens, but valuable when the fast teammate disappears. This is an action--turnover interaction. A standard shared advantage observes only the final return. A purely event-conditioned critic predicts the drop more accurately but may still provide an unclear actor signal. TOCA assigns the unavoidable replacement penalty to $q_E$ and the useful compensatory movement to $q_{AE}$.

The same logic applies beyond navigation: a sensing robot can reposition after a sensor failure, or a human--AI assistant can summarize context so that an incoming participant can contribute quickly. In each case, the membership change itself is not the survivor's action, but preparation for or response to that change is.

The central assumption for turnover-orthogonal credit is \emph{exogenous turnover}:
\begin{equation}
E_t\perp A_t\mid H_t
\label{eq:exo}
\end{equation}
This assumption does not mean that turnover is unimportant. It means that, conditioned on the current history, the event is not caused by the sampled joint action. This covers randomized replacement, device failure, externally scheduled team changes, and stress-test evaluations. If turnover is action-caused, then event effects may legitimately be part of action credit;
as discussed in the \emph{Additional Analysis and Limitations} section.

The assumption also clarifies the role of the state. We do not require events to be independent of the history. Replacement may be more likely in difficult states or after a fixed schedule. What we require is that, after conditioning on the history used by the critic, the sampled action is not the cause of the event. Under this condition, subtracting an event-conditioned but action-independent baseline does not bias the policy gradient, while it can reduce misattribution and variance.

\section{Turnover-Orthogonal Decomposition}

\subsection{Credit-Faithfulness Criterion}

We say that an advantage estimator is turnover-faithful if adding a pure event offset to the return does not change the expected actor update. Formally, let $c(h,e)$ be any scalar that depends on the history and event but not on the sampled joint action. For an advantage estimator $\widehat A$, turnover faithfulness requires that replacing the return target by $G_t+c(h,e)$ changes only the baseline term and not the action-dependent policy-gradient direction. An ordinary shared return target does not satisfy this property unless its baseline also tracks $c(h,e)$. The event-conditioned baseline in Equation~\ref{eq:baseline} is the minimal baseline that removes such offsets while retaining the interaction residual.

This criterion is weaker than demanding perfect causal identification in the full environment. It is intended for the common experimental and operational setting where turnover events are externally injected or logged. In that setting, the method does not decide whether a replacement was good or bad; it decides whether the resulting change should be assigned to the agents' actions.

\subsection{Centered Value Decomposition}

For a fixed history $h$, consider the random variables $A\sim \pi(\cdot\mid h)$ and $E\sim p(\cdot\mid h)$. Under exogenous turnover, the joint distribution factorizes as $p(a,e\mid h)=\pi(a\mid h)p(e\mid h)$. We decompose $Q^{\pi}(h,a,e)$ as in Equation~\ref{eq:decomp} with centering constraints
\begin{align}
\mathbb E_A[q_A(h,A)]&=0, & \mathbb E_E[q_E(h,E)]&=0,\label{eq:center1}\\
\mathbb E_A[q_{AE}(h,A,e)]&=0, & \mathbb E_E[q_{AE}(h,a,E)]&=0 \label{eq:center2}
\end{align}
The decomposition mirrors an analysis-of-variance decomposition for an event-conditioned value function. The terms are identifiable up to function approximation error when expectations are taken under the policy and event distribution:
\begin{align}
\mu(h)&=\mathbb E_{A,E}[Q^{\pi}(h,A,E)]\\
q_A(h,a)&=\mathbb E_E[Q^{\pi}(h,a,E)]-\mu(h)\\
q_E(h,e)&=\mathbb E_A[Q^{\pi}(h,A,e)]-\mu(h)\\
q_{AE}(h,a,e)&=Q^{\pi}(h,a,e)-\mu(h)-q_A(h,a)-q_E(h,e)
\end{align}
Thus $q_E$ captures value changes due only to turnover, while $q_{AE}$ captures whether the action is more or less useful under a particular event.

\begin{table}[t]
\centering
\small
\begin{tabular}{ll}
\toprule
Term & Credit interpretation \\
\midrule
$\mu(h)$ & expected value of the current history \\
$q_A(h,a)$ & action effect averaged over events \\
$q_E(h,e)$ & pure turnover effect averaged over actions \\
$q_{AE}(h,a,e)$ & action-specific robustness or fragility to event \\
\bottomrule
\end{tabular}
\caption{Interpretation of TOCA components. Only $q_A$ and $q_{AE}$ should affect the actor update.}
\label{tab:components}
\end{table}

\textbf{Proposition 1:} Under Equation~\ref{eq:exo}, using $B^{\pi}(h,e)=\mu(h)+q_E(h,e)$ as a policy-gradient baseline removes all additive pure-turnover offsets without changing the expected policy-gradient direction.

\textbf{Proof sketch:} Let $c(h,e)$ be any event-dependent, action-independent offset. Its policy-gradient contribution is $\mathbb E[\nabla_\theta\log\pi_\theta(a\mid h)c(h,e)]$. Here the score function is $\nabla_\theta\log\pi_\theta(a\mid h)$, whose expectation under $a\sim\pi_\theta(\cdot\mid h)$ is zero. By exogeneity, $c(h,e)$ separates from the action score, so the pure event effect is a valid baseline term. The interaction term does not factor this way because it depends jointly on $a$ and $e$, and removing it would discard valid action-dependent credit.

\subsection{Turnover-Orthogonal Advantage}

The event-conditioned baseline is
\begin{equation}
B^{\pi}(h,e)=\mathbb E_{A\sim\pi(\cdot\mid h)}Q^{\pi}(h,A,e)=\mu(h)+q_E(h,e) 
\label{eq:baseline}
\end{equation}
It depends on the turnover event but not on the sampled action. The turnover-orthogonal advantage is
\begin{equation}
A^{\mathrm{TO}}(h,a,e)=Q^{\pi}(h,a,e)-B^{\pi}(h,e)\\=q_A(h,a)+q_{AE}(h,a,e)
\label{eq:adv}
\end{equation}
Equation~\ref{eq:adv} removes pure event credit $q_E$ while retaining the action--event interaction. This distinction is important. If an agent is unlucky because a capable teammate is replaced, the pure replacement penalty should not be assigned to that agent. If the same agent previously moved to cover a landmark that a replacement is unlikely to cover, the resulting interaction should remain part of its credit.

For per-agent counterfactual credit, define
\begin{equation}
B_i^{\mathrm{TO}}(h,a_{-i},e)=\mathbb E_{a_i'\sim \pi_i(\cdot\mid o_i)}Q^{\pi}(h,(a_i',a_{-i}),e)
\end{equation}
and
\begin{equation}
A_i^{\mathrm{TO}}=Q^{\pi}(h,a,e)-B_i^{\mathrm{TO}}(h,a_{-i},e)
\end{equation}
This compares agent $i$'s action to alternatives under the same turnover event. In experiments, we use a shared actor loss with either the event-conditioned baseline in Equation~\ref{eq:baseline} or its softly weighted variant below.

\subsection{Soft Interaction Credit}

In high-variance continuous-control-like tasks, a learned interaction head may be noisy early in training. We therefore introduce TOCA-$\beta$:
\begin{equation}
A^{\mathrm{TO}\text{-}\beta}(h,a,e)=q_A(h,a)+\beta q_{AE}(h,a,e),\quad \beta\in[0,1]
\label{eq:beta}
\end{equation}
The special case $\beta=0$ removes interaction credit; $\beta=1$ recovers full TOCA. Equivalently, the actor subtracts $\mu+q_E+(1-\beta)q_{AE}$ from the target. We use $\beta=0.5$ for Dynamic Spread after a small ablation over $\{0,0.1,0.3,0.5,0.7,1\}$.

\section{Method}

\subsection{Permutation-Invariant Critic}

The critic receives a variable-size set of active-agent features and an event token. Each active agent contributes a feature vector containing its observation, action, active mask, and optional capability. We use shared encoders followed by pooling to obtain a state representation $g_s$ and a state-action representation $g_{sa}$. A key implementation detail is that the pure baseline terms must be action-independent. Here $\phi$ denotes the learnable parameters of the centralized critic,
including the shared set encoders and the four output heads
$f_\mu$, $f_A$, $f_E$, and $f_{AE}$:
\begin{align}
\mu_\phi(h)&=f_\mu(g_s), & q_{E,\phi}(h,e)&=f_E(g_s,e),\\
q_{A,\phi}(h,a)&=f_A(g_{sa}), & q_{AE,\phi}(h,a,e)&=f_{AE}(g_{sa},e)
\end{align}
Using state-action features for $\mu$ or $q_E$ would make the baseline action-dependent and would violate the orthogonality intended by Equation~\ref{eq:baseline}. The full critic is
\begin{equation}
Q_\phi(h,a,e)=\mu_\phi(h)+q_{A,\phi}(h,a)+q_{E,\phi}(h,e)+q_{AE,\phi}(h,a,e)
\end{equation}
The critic is permutation-invariant over active agents because each agent is encoded with shared weights and then pooled over the active set.

\subsection{Training Objective}

The critic is trained by squared Bellman error together with centering regularizers. For a target $y_t=r_t+\gamma V_{\bar\phi}(h_{t+1},e_{t+1})$, the value loss is
\begin{equation}
\mathcal L_Q=\mathbb E[(Q_\phi(h_t,a_t,e_t)-y_t)^2]
\end{equation}
The centering loss enforces approximate versions of Equations~\ref{eq:center1}--\ref{eq:center2} using minibatch averages and counterfactual samples:
\begin{equation}
\mathcal L_{\mathrm{ctr}}=\|\widehat{\mathbb E}_A q_A\|^2+\|\widehat{\mathbb E}_E q_E\|^2+\|\widehat{\mathbb E}_A q_{AE}\|^2+\|\widehat{\mathbb E}_E q_{AE}\|^2
\end{equation}
The actor uses a PPO-style clipped objective \cite{schulman2017ppo} with $A^{\mathrm{TO}}$ or $A^{\mathrm{TO}\text{-}\beta}$. The total loss is
\begin{equation}
\mathcal L=\mathcal L_{\mathrm{actor}}+c_v\mathcal L_Q+c_c\mathcal L_{\mathrm{ctr}}-c_H\mathcal H(\pi)
\end{equation}
where $\mathcal H$ is policy entropy.

\subsection{Practical Estimation}

In implementation, the expectations in the centering constraints are approximated with minibatch averages over sampled actions and events. For discrete actions, counterfactual action samples can be enumerated for small action spaces or sampled from the current policy. For the replacement event, we sample event tokens from the empirical event distribution in the replayed batch. This produces a lightweight regularizer rather than a separate supervised objective for each component.

The critic uses two information streams: The state-only stream encodes active positions, capabilities, masks, and event-independent features. The state-action stream additionally appends one-hot or embedded actions. The separation is crucial: $\mu$ and $q_E$ are produced only from the state-only stream, while $q_A$ and $q_{AE}$ use the state-action stream. Early experiments with a single state-action stream made the baseline action-dependent and degraded Dynamic Spread performance. This observation is consistent with the theory, because an action-dependent baseline can subtract away the very policy signal it is meant to preserve.

TOCA adds little conceptual overhead to MAPPO-style training. The actor and rollout procedure are unchanged. The additional cost is in the critic heads and the centering loss. For a pooled set representation of dimension $d$, the decomposition heads are small multilayer perceptrons. Thus, the asymptotic cost is dominated by the same set encoder used by the centralized critic. In the experiments below, this overhead was small compared with environment simulation.

\subsection{Metrics}

We use three metric families: average episode return for performance, high-event return for robustness at the largest replacement rate, and diagnostic faithfulness when ground-truth components are available. For $j\in\{A,E,AE\}$, component error is
\begin{equation}
E_{q_j}=\frac{1}{|\mathcal D|}\sum_{(h,a,e)\in\mathcal D}\left|\hat q_j(h,a,e)-q_j^\star(h,a,e)\right|
\end{equation}
with unused arguments suppressed for $q_A$ and $q_E$. Let $\mathcal S=\{t:e_t\neq 0\}$ be diagnostic turnover-shock steps, $C_t^\star=q_A^\star(h_t,a_t)+q_{AE}^\star(h_t,a_t,e_t)$, and $\hat C_t=\hat q_A(h_t,a_t)+\hat q_{AE}(h_t,a_t,e_t)$. Here $(\cdot)^\star$ denotes the ground-truth diagnostic component,
while $\hat{(\cdot)}$ denotes the corresponding learned critic estimate.

We report
\begin{equation}
\mathrm{ShockMAE}=\frac{1}{|\mathcal S|}\sum_{t\in\mathcal S}|\hat C_t-C_t^\star|
\end{equation}
which measures action credit assigned around exogenous turnover after removing the pure event term. Lower values indicate more faithful credit.

\begin{algorithm}[t]
\caption{Turnover-Orthogonal Actor-Critic Update}
\label{alg:toca}
\begin{algorithmic}[1]
\STATE Collect trajectories with active sets $\mathcal N_t$, actions $a_t$, rewards $r_t$, and events $e_t$.
\STATE Encode active-agent set to obtain $g_s$ and state-action set to obtain $g_{sa}$.
\STATE Predict $\mu_\phi$, $q_{A,\phi}$, $q_{E,\phi}$, and $q_{AE,\phi}$.
\STATE Train $Q_\phi=\mu+q_A+q_E+q_{AE}$ with Bellman loss and centering loss.
\STATE Form actor advantage $A=q_A+\beta q_{AE}$, with $\beta=1$ for full TOCA.
\STATE Update decentralized actor using PPO-style clipping and entropy regularization.
\end{algorithmic}
\end{algorithm}

\section{Experiments}

We evaluate two questions: First, does the decomposition recover useful credit signals when the ground-truth sources of return variation are controlled? Second, does turnover-orthogonal credit improve robustness in a MARL control benchmark with replacement events?

\subsection{Diagnostic Environment}

The diagnostic environment is a controlled open-team task with known ground-truth components $q_A^\star$, $q_E^\star$, and $q_{AE}^\star$. The return combines additive action and event effects with a multiplicative action--event interaction; event effects vary with replacement rate and capability shock, while the interaction rewards actions useful only under particular turnover events.

This benchmark acts as a unit test for credit assignment rather than a full physics task. A method can obtain high return while placing event effects in the action head, so we report both policy return and decomposition faithfulness. We compare MAPPO, an event-aware set critic, TOCA, TOCA without interaction, and TOCA without centering over five seeds, team sizes $N\in\{3,4,5\}$, and event rates $p\in\{0,0.05,0.1,0.2\}$, yielding 360 completed runs.

\begin{table*}[t]
\centering
\small
\begin{tabular}{lc}
\toprule
\multicolumn{2}{c}{(a) Policy performance on the diagnostic benchmark} \\
\toprule
Method & Avg. Return $\uparrow$ \\
\midrule
MAPPO & $319.375 \pm 1.267$ \\
Event-Set-MAPPO & $319.419 \pm 1.271$ \\
TOCA & $341.436 \pm 6.928$ \\
TOCA w/o interaction & $299.437 \pm 8.817$ \\
TOCA w/o centering & $374.070 \pm 3.422$ \\
\bottomrule
\end{tabular}

\vspace{0.45em}

\begin{tabular}{lcccc}
\toprule
\multicolumn{5}{c}{(b) Decomposition quality for methods with explicit TOCA heads} \\
\toprule
Method & $E_{q_A}\downarrow$ & $E_{q_E}\downarrow$ & $E_{q_{AE}}\downarrow$ & Shock MAE $\downarrow$ \\
\midrule
TOCA & $3.486 \pm 0.004$ & $2.798 \pm 0.019$ & $1.619 \pm 0.018$ & $2.135 \pm 0.015$ \\
TOCA w/o interaction & $3.503 \pm 0.002$ & $2.743 \pm 0.025$ & $1.474 \pm 0.025$ & $2.164 \pm 0.015$ \\
TOCA w/o centering & $0.705 \pm 0.114$ & $0.902 \pm 0.056$ & $3.065 \pm 0.066$ & $3.629 \pm 0.114$ \\
\bottomrule
\end{tabular}
\caption{Diagnostic benchmark split into policy performance and decomposition quality. Component errors and shock misattribution are reported only for methods that learn explicit TOCA decomposition heads; MAPPO-style baselines do not define $q_A$, $q_E$, or $q_{AE}$.}
\label{tab:diagnostic}
\end{table*}

Table~\ref{tab:diagnostic} separates policy performance from decomposition quality. Panel (a) shows that TOCA improves average return over MAPPO and Event-Set-MAPPO. Removing the interaction term reduces return substantially, from $341.436$ to $299.437$, indicating that action--turnover interaction credit is important in the controlled setting. Panel (b) reports component errors only for methods with explicit decomposition heads. The no-centering ablation achieves the highest raw return, but its interaction error and shock misattribution are much worse. We interpret this as a return--faithfulness tradeoff: without centering, the network can fit returns while mixing the semantic roles of the components. Since the goal of TOCA is credit-faithful decomposition, centering is essential.

\subsection{Dynamic Spread with Replacement Events}

We next evaluate a replacement-only Dynamic Spread benchmark based on cooperative particle navigation. Four active agents must cover landmarks. At each event, a high-capability active agent is replaced by a weak agent while the active team size remains fixed. This isolates turnover from team-size changes and makes the event disruptive: a greedy sanity policy degrades from $-7.24$ at event rate $0$ to $-28.64$ at event rate $0.3$.

The reward is the negative average distance from landmarks to their closest effective covering agent, plus collision penalties. Replacement affects the effective coverage distance by lowering the capability of the incoming agent. This makes replacement harmful even when the new agent inherits the old position. Thus, the event is not merely a random perturbation; it changes the team's ability to cover landmarks. We use replacement rather than addition or removal because it keeps the active team size fixed and makes the comparison focused on credit assignment rather than variable action-space size.

The high-event setting is particularly important. At event rate zero, turnover-aware structure is less useful because the task reduces to an ordinary fixed-team navigation problem. At high event rates, the policy repeatedly observes that teammates change capability, making it valuable to learn which actions are robust to replacement. This is the regime where TOCA should help.

We train MAPPO, Event-Set-MAPPO, TOCA-$\beta$ with $\beta=0.5$, and TOCA without interaction. We use five seeds and two training event rates, $0.1$ and $0.2$, for 40 runs. Evaluation uses event rates $\{0,0.1,0.2,0.3\}$. Returns are negative distances and penalties, so higher is better.

\begin{table}[t]
\centering
\small
\begin{tabular}{lcc}
\toprule
Method & Avg. Return $\uparrow$ & High-Event Return $\uparrow$ \\
\midrule
MAPPO & $-32.166 \pm 0.638$ & $-39.300$ \\
Event-Set-MAPPO & $-32.090 \pm 0.717$ & $-39.214$ \\
TOCA-$\beta$ ($\beta=0.5$) & $\mathbf{-32.063 \pm 0.724}$ & $\mathbf{-38.329}$ \\
TOCA w/o interaction & $-32.449 \pm 0.663$ & $-41.369$ \\
\bottomrule
\end{tabular}
\caption{Replacement-only Dynamic Spread. TOCA-$\beta$ is competitive in average return and achieves the best high-turnover return.}
\label{tab:spreadmain}
\end{table}

Table~\ref{tab:spreadmain} shows that TOCA-$\beta$ obtains the best mean return, although the margin over Event-Set-MAPPO is modest. The high-event return is more informative: TOCA-$\beta$ improves over Event-Set-MAPPO and substantially improves over the no-interaction ablation. This suggests that softly weighted interaction credit helps agents remain robust when replacements are frequent.

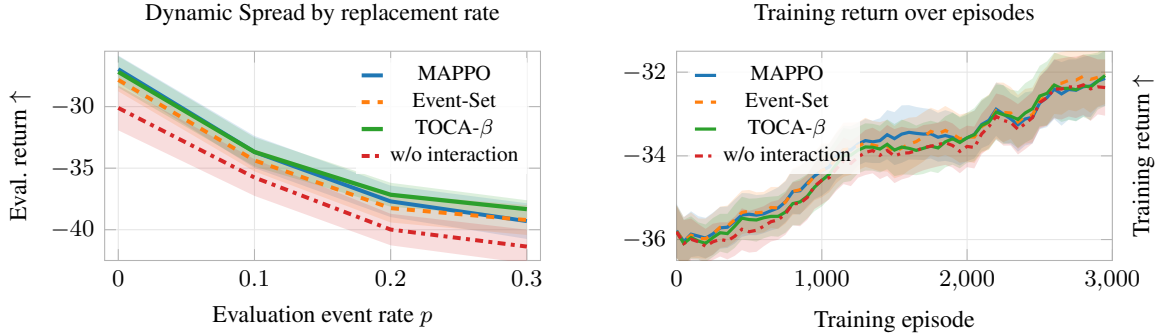
\begin{figure*}[t]
\centering
\begin{tikzpicture}
\begin{groupplot}[
    group style={group size=2 by 1, horizontal sep=1.8cm},
    width=0.445\textwidth,
    height=4.35cm,
    tick label style={font=\footnotesize},
    label style={font=\footnotesize},
    title style={font=\footnotesize},
    legend style={font=\scriptsize, draw=none, fill=white, fill opacity=0.85, text opacity=1, inner xsep=1.2pt, inner ysep=1.0pt, row sep=-1pt},
    legend image code/.code={\draw[#1] (0cm,0cm) -- (0.28cm,0cm);},
    ymajorgrids,
    xmajorgrids,
    grid style={gray!20},
    axis line style={draw=gray!60},
    tick style={draw=gray!60},
]

\nextgroupplot[
    title={Dynamic Spread by replacement rate},
    xlabel={Evaluation event rate $p$},
    ylabel={Eval. return $\uparrow$},
    xmin=-0.01,
    xmax=0.31,
    ymin=-42.5,
    ymax=-25.5,
    xtick={0,0.1,0.2,0.3},
    legend pos=north east,
]

\addplot[name path=mappoupper, draw=none, forget plot] coordinates {(0,-25.846) (0.1,-32.376) (0.2,-36.419) (0.3,-37.849)};
\addplot[name path=mappolower, draw=none, forget plot] coordinates {(0,-28.074) (0.1,-34.989) (0.2,-38.977) (0.3,-40.752)};
\addplot[mappoblue, fill opacity=0.16, draw=none, forget plot] fill between[of=mappoupper and mappolower];
\addplot[color=mappoblue, line width=1.5pt, mark=none] coordinates {(0,-26.960) (0.1,-33.683) (0.2,-37.698) (0.3,-39.300)};
\addlegendentry{MAPPO}

\addplot[name path=eventupper, draw=none, forget plot] coordinates {(0,-26.892) (0.1,-33.467) (0.2,-37.079) (0.3,-38.003)};
\addplot[name path=eventlower, draw=none, forget plot] coordinates {(0,-28.762) (0.1,-35.265) (0.2,-39.413) (0.3,-40.425)};
\addplot[eventorange, fill opacity=0.16, draw=none, forget plot] fill between[of=eventupper and eventlower];
\addplot[color=eventorange, line width=1.5pt, mark=none, dashed] coordinates {(0,-27.827) (0.1,-34.366) (0.2,-38.246) (0.3,-39.214)};
\addlegendentry{Event-Set}

\addplot[name path=tocaupper, draw=none, forget plot] coordinates {(0,-25.907) (0.1,-32.495) (0.2,-36.186) (0.3,-37.607)};
\addplot[name path=tocalower, draw=none, forget plot] coordinates {(0,-28.455) (0.1,-34.871) (0.2,-38.133) (0.3,-39.050)};
\addplot[tocagreen, fill opacity=0.16, draw=none, forget plot] fill between[of=tocaupper and tocalower];
\addplot[color=tocagreen, line width=1.6pt, mark=none] coordinates {(0,-27.181) (0.1,-33.683) (0.2,-37.159) (0.3,-38.329)};
\addlegendentry{TOCA-$\beta$}

\addplot[name path=nointupper, draw=none, forget plot] coordinates {(0,-28.287) (0.1,-34.266) (0.2,-38.705) (0.3,-39.997)};
\addplot[name path=nointlower, draw=none, forget plot] coordinates {(0,-31.924) (0.1,-37.250) (0.2,-41.269) (0.3,-42.741)};
\addplot[ablred, fill opacity=0.16, draw=none, forget plot] fill between[of=nointupper and nointlower];
\addplot[color=ablred, line width=1.5pt, mark=none, dashdotted] coordinates {(0,-30.106) (0.1,-35.758) (0.2,-39.987) (0.3,-41.369)};
\addlegendentry{w/o interaction}

\nextgroupplot[
    title={Training return over episodes},
    xlabel={Training episode},
    ylabel={Training return $\uparrow$},
    ylabel style={at={(axis description cs:1.12,0.5)},anchor=south},
    xmin=0,
    xmax=3000,
    ymin=-36.5,
    ymax=-31.5,
    xtick={0,1000,2000,3000},
    legend pos=north west,
]

\addplot[name path=mappotrainupper, draw=none, forget plot] table[x=episode,y=upper,col sep=comma] {lc_mappo.csv};
\addplot[name path=mappotrainlower, draw=none, forget plot] table[x=episode,y=lower,col sep=comma] {lc_mappo.csv};
\addplot[mappoblue, fill opacity=0.16, draw=none, forget plot] fill between[of=mappotrainupper and mappotrainlower];
\addplot+[color=mappoblue, very thick, mark=none] table[x=episode,y=mean,col sep=comma] {lc_mappo.csv};
\addlegendentry{MAPPO}

\addplot[name path=eventtrainupper, draw=none, forget plot] table[x=episode,y=upper,col sep=comma] {lc_event_set.csv};
\addplot[name path=eventtrainlower, draw=none, forget plot] table[x=episode,y=lower,col sep=comma] {lc_event_set.csv};
\addplot[eventorange, fill opacity=0.16, draw=none, forget plot] fill between[of=eventtrainupper and eventtrainlower];
\addplot+[color=eventorange, very thick, dashed, mark=none] table[x=episode,y=mean,col sep=comma] {lc_event_set.csv};
\addlegendentry{Event-Set}

\addplot[name path=tocatrainupper, draw=none, forget plot] table[x=episode,y=upper,col sep=comma] {lc_toca_beta.csv};
\addplot[name path=tocatrainlower, draw=none, forget plot] table[x=episode,y=lower,col sep=comma] {lc_toca_beta.csv};
\addplot[tocagreen, fill opacity=0.16, draw=none, forget plot] fill between[of=tocatrainupper and tocatrainlower];
\addplot+[color=tocagreen, very thick, mark=none] table[x=episode,y=mean,col sep=comma] {lc_toca_beta.csv};
\addlegendentry{TOCA-$\beta$}

\addplot[name path=nointtrainupper, draw=none, forget plot] table[x=episode,y=upper,col sep=comma] {lc_no_interaction.csv};
\addplot[name path=nointtrainlower, draw=none, forget plot] table[x=episode,y=lower,col sep=comma] {lc_no_interaction.csv};
\addplot[ablred, fill opacity=0.16, draw=none, forget plot] fill between[of=nointtrainupper and nointtrainlower];
\addplot+[color=ablred, very thick, dashdotted, mark=none] table[x=episode,y=mean,col sep=comma] {lc_no_interaction.csv};
\addlegendentry{w/o interaction}

\end{groupplot}
\end{tikzpicture}
\caption{Dynamic Spread return plots. Left: evaluation return versus replacement event rate, shown with mean lines and shaded standard-error bands for a cleaner comparison. Right: training return versus episode, averaged over five seeds and two training event rates. TOCA-$\beta$ is most useful at high turnover and learns competitively with the event-aware baseline.}
\label{fig:pgfplots}
\end{figure*}

Figure~\ref{fig:pgfplots} shows both evaluation robustness and learning dynamics for Dynamic Spread. At low event rates, MAPPO is best or tied. At high event rates, TOCA-$\beta$ becomes best. The learning curves show that the methods have similar early training behavior, while TOCA-$\beta$ remains competitive through the end of training. This matches the motivation for turnover-orthogonal credit: when replacements are rare, ordinary action credit is often sufficient; when replacements are frequent, removing pure turnover effects and preserving action--turnover interactions becomes more useful.

Matched-condition paired comparisons give the same picture. Across 40 matched evaluation conditions, TOCA-$\beta$ improves over Event-Set-MAPPO by $0.825\pm0.695$ return with a $67.5\%$ win rate, and improves over TOCA without interaction by $2.717\pm0.869$. Restricting to high-turnover evaluations $p\geq0.2$, the corresponding improvements are $0.986\pm1.006$ and $2.934\pm1.051$. These results suggest that TOCA-$\beta$ provides a consistent robustness trend over Event-Set-MAPPO under frequent replacement, while the clearer effect is its improvement over the no-interaction ablation, which directly supports the importance of the action--turnover interaction term $q_{AE}$.

\subsection{Qualitative Interpretation of Results}

The diagnostic and Dynamic Spread experiments serve complementary roles. The diagnostic benchmark verifies the decomposition directly: full TOCA outperforms the no-interaction variant, showing that $q_{AE}$ can be learned and can improve optimization when interaction structure is clean. Dynamic Spread is harder because return depends on geometry, collisions, and replacement-induced capability changes. In that setting, full interaction credit can be noisy, but the softened TOCA-$\beta$ signal is most useful at high event rates.

This pattern supports the main design principle: Pure turnover should be removed as a baseline, while action--turnover interaction should not be discarded. Event-Set-MAPPO is a strong baseline because it observes replacement tokens and predicts event-conditioned returns, but it does not specify which event-conditioned variations are action credit. TOCA-$\beta$ helps where this distinction matters most: high replacement rates and comparisons against the no-interaction ablation.

\subsection{Beta Ablation}

The beta sweep compares $\beta\in\{0,0.1,0.3,0.5,0.7,1\}$. We select $\beta=0.5$ because it gives the strongest high-event return among stable intermediate settings. The ablation suggests that interaction credit is useful, but injecting it at full strength can add variance in noisy control benchmarks.

\section{Additional Analysis and Limitations}
\label{sec:limitations}

TOCA changes the actor signal from ``what return followed this transition?'' to ``what return followed after accounting for the turnover event?'' This matters when actions can change the team's sensitivity to turnover: an agent may duplicate a teammate's coverage, hedge against replacement, or communicate state that helps an incoming teammate. If events only add noise and agents cannot respond to them, then $q_{AE}$ should be small and the no-interaction variant should match TOCA-$\beta$.

The framework is designed for exogenous or externally specified turnover. If the event itself is caused by an agent's action, removing $q_E$ may remove legitimate action credit; endogenous turnover therefore requires explicit causal modeling. A second limitation is statistical strength: Dynamic Spread is best interpreted as a robustness trend rather than a decisive benchmark victory over Event-Set-MAPPO. The strongest evidence is the diagnostic decomposition and the paired improvement over the no-interaction ablation under high turnover.

\section{Broader Applicability}

Although the experiments focus on cooperative navigation-style tasks, the decomposition is domain-agnostic. Any open-team task with logged joins, departures, failures, or replacements can form event tokens and train the same critic heads. Suitable application classes include robot fleets, warehouse systems, sensor networks, service teams, and human--AI collaboration, where evaluating only average return can hide whether a policy remains effective under unexpected team change. For deployments involving human participants, event descriptors should avoid sensitive or protected attributes; TOCA is a cooperative-control credit assignment method, not a personnel evaluation mechanism.

\section{Conclusion}

We introduced TOCA, a turnover-orthogonal credit assignment framework for open-team MARL. TOCA decomposes event-conditioned value into action, turnover, and interaction terms, removing pure turnover credit while preserving credit for actions that help the team respond to replacement events. Diagnostic experiments validate the decomposition and show that interaction credit matters. Dynamic Spread experiments show that TOCA-$\beta$ improves high-turnover robustness and outperforms the no-interaction ablation. These results suggest that separating action credit from population-turnover credit is a promising direction for robust learning in dynamic cooperative teams.

\end{document}